\documentclass[journal]{IEEEtran}

\usepackage{graphicx}

\begin{document}
%
\title{Bandwidth and dynamic range of a widely tunable Josephson parametric amplifier}
%
%
%


\author{M.~A.~Castellanos-Beltran, K.~D.~Irwin, L.~R.~Vale, G.~C.~Hilton, and K.~W.~Lehnert

\thanks{M.~A.~Castellanos-Beltran is with JILA and the Department of Physics, the University of
Colorado at Boulder, Boulder, CO 80309-0440 USA.}
\thanks{K.~W.~Lehnert is with JILA and the Department of Physics, the
University of Colorado at Boulder, Boulder, Colorado, 80309-0440,
and also with the National Institute of Standards and Technology,
Boulder, CO 80305-3328 USA (e-mail:
konrad.lehnert@jila.colorado.edu).}
\thanks{K.~D.~Irwin, G.~C.~Hilton and L. R. Vale are
with the National Institute of Standards and Technology, Boulder, CO
80305-3328 USA. } }

%
%




\maketitle

\begin{abstract}

We characterize the signal bandwidth and dynamic range of a recently
developed type of Josephson parametric amplifier. These amplifiers
consist of a series array of SQUIDs embedded in a microwave cavity.
They are narrow band, only amplifying signals close to the cavity's
resonance frequency, but the cavity's resonance frequency, and hence
the amplified band, can be widely tuned. For a particular
realization of these amplifiers we measure how the signal bandwidth
depends on amplifier's gain. We find that the amplitude gain times
signal bandwidth is approximately the linewidth of the cavity. In
addition we measure the amplifier's dynamic range and saturation
power.

\end{abstract}



\IEEEpeerreviewmaketitle

\section{Introduction}

\IEEEPARstart{T}{he} ability to manipulate quantum information
encoded in microwave fields has led to a renewed interest in
Josephson parametric amplifiers (JPAs) \cite{Castellanos,
yamamoto:042510,devoret}. For these applications the ability of JPAs
to amplify signals with the least amount of added noise is critical
\cite{caves1,yurke1988,yurke1989,yurke1990}. Unfortunately JPAs are
typically narrow band amplifiers with small dynamic range. It is
therefore important to understand the bandwidth and dynamic range of
any particular JPA in order to determine if it is appropriate for
these applications.

We recently introduced a new kind of JPA. Although it is still
narrow band, the amplified band can be tuned over a full octave
\cite{Castellanos}. We have shown that it has good noise performance
and can squeeze the vacuum noise by 10 dB \cite{Castellanos2}. Here
we characterize other important parameters of this amplifier,
specifically the signal-bandwidth, dynamic range and saturation
power.


\begin{figure}
\includegraphics{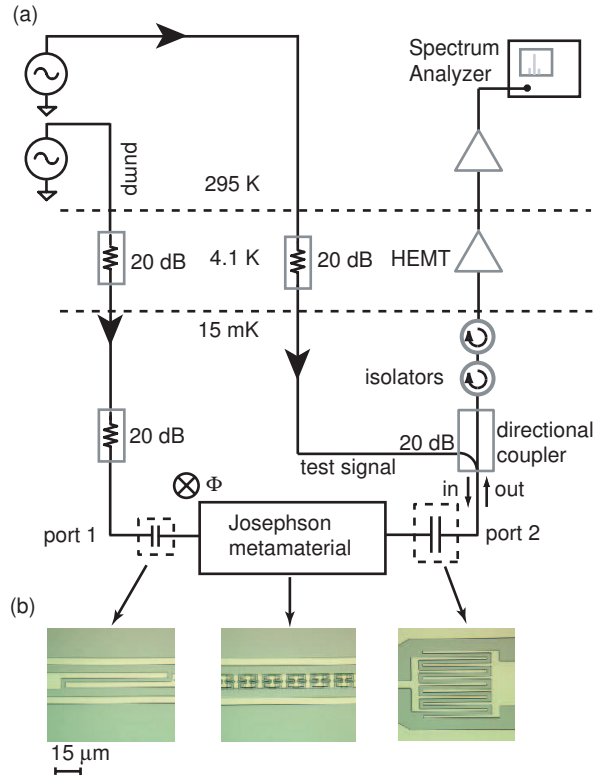}
\caption{\label{fig:Fig1} (a) Measurement schematic. The device is
cooled down to 15~mK using a dilution cryostat. Two microwave
generators are used to study the JPA: one creates the pump tone,
while the second creates a signal tone. The pump is injected through
the weakly coupled port (port 1), while the signal tone is incident
on the strongly coupled port (port 2). Signals emerging from port 2
go through a pair of isolators before being amplified by a cryogenic
high-electron-mobility transistor amplifier (HEMT) and a room
temperature amplifier. (b) Pictures of the device.}
\end{figure}

\section{Josephson metamaterial parametric amplifier}\label{sec:JM}

Our realization of a JPA consists of a length of non-linear
transmission line from which a half-wavelength cavity is made. The
nonlinear transmission line is a coplanar waveguide transmission
line where the center conductor is a series array of
SQUIDs\cite{Castellanos}. The inductance per unit length of this
transmission line comes mostly from the nonlinear Josephson
inductance of the SQUIDs and not from the geometrical inductance.
Furthermore, the inductance per length is tunable with magnetic flux
applied to the SQUIDs loops. As a consequence, the phase velocity of
waves in this transmission line depends both on intensity of those
waves and on the applied flux. At microwave frequencies the SQUIDs
are spaced much less than a wavelength apart, so we treat the array
as a continuous medium with flux tunable and intensity dependent
phase velocity (c.f. an optical Kerr medium). We call this nonlinear
effective medium a Josephson metamaterial (Fig. \ref{fig:Fig1}).

We create a half-wavelength microwave cavity from this metamaterial
by interrupting the center conductor with two capacitors, one much
bigger than the other, thus creating a strongly coupled port and a
weakly coupled port (Fig. \ref{fig:Fig1}). Because the resonance
frequency ($f_{res}$) of this cavity is proportional to the phase
velocity, it is dependant on the intensity of the microwave field
stored in the cavity. When a large tone, referred to as the pump, is
applied through the weakly coupled port close to the resonance
frequency of the cavity, it makes the resonance frequency oscillate
at twice the pump frequency. This process creates parametric
gain\cite{Louisell2}.

\section{Design and fabrication}

\begin{figure}
\includegraphics{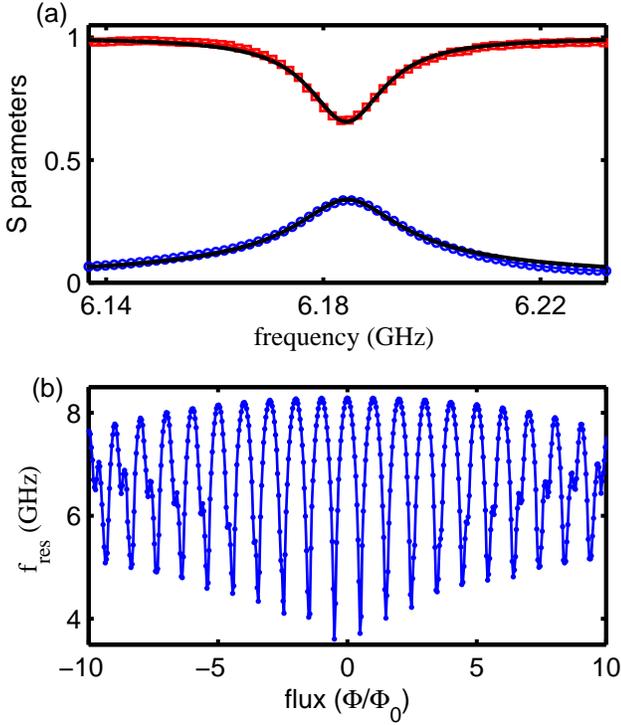}
\caption{\label{fig:Fig2} (a) Magnitude of the transmission
$|S_{21}|$ (circles) and reflection $|S_{22}|$ (squares)
coefficients as a function of frequency at $\Phi=0.37\Phi_0$. From
the fits (lines), we extract the resonance frequency $f_{res}=6.184
$~GHz, and the amplitude decay rates associated with port 1, port 2
and internal loss $\gamma_{c1}=2\pi \times 323$~kHz,
$\gamma_{c2}=2\pi \times 7.805$~MHz, and $\gamma_{i}=2\pi \times
1.277$~MHz, respectively. (b) Resonance frequency as a function of
magnetic flux (points connected by lines)}
\end{figure}

Devices were fabricated at NIST Boulder on a high purity
uncompensated silicon wafer with resistivity $> 17$ k$\Omega$cm. A
standard Nb/AlOx/Nb trilayer process \cite{NISTfab} was used,
modified by eliminating the shunt resistor layer and minimizing
deposited oxides \cite{Mates}. Some of the features of this sample
were already explained in a previous letter \cite{Castellanos2}, so
we will only mention the measured parameters. The SQUIDs have an
average critical current $I_c$ per SQUID in the JPA of 31~$\mu$A.
This value is close to the designed value of 30~$\mu$A. The coplanar
waveguide transmission line was designed to have a capacitance per
unit length of $C_l=0.15$ nF/m and a geometrical inductance per unit
length of $L_l=0.49$ $\mu$H/m. The coupling capacitances were
estimated to be 4.6 fF and 22.6 fF using microwave simulations.
Based on measurements of the S-parameters for low powers (Fig.
\ref{fig:Fig2}a), these values are very close to the observed ones.
From the expected impedance and phase velocity of the transmission
line as well as the coupling capacitors' values, we predicted a
half-wave resonance frequency of $8.23$ GHz with no applied flux.
This value agrees with the measured resonance frequency within 3\%
(Fig. \ref{fig:Fig2}b).

In Fig. \ref{fig:Fig2}b we show how the flux tunes the cavity's
resonance frequency over many periods. Such a plot provides a
qualitative measure of the uniformity of the SQUIDs that comprise
the device. In particular, the uniformity is much better than the
device in \cite{Castellanos} which was fabricated using electron
beam lithography and double angle evaporation.



\section{Operation}

The JPA is operated in four photon mode: as described in section
\ref{sec:JM}, the pump tone, with frequency $f_p$, is applied
through the weakly coupled port close to the resonance frequency of
the cavity. The large pump makes the resonance frequency oscillate
at twice the pump frequency, creating parametric gain
\cite{Louisell2}. The JPA operates as a reflection amplifier as
shown in Fig. \ref{fig:Fig1}(b). When the signal with frequency
$f_s$ is detuned from the pump, the parametric amplifier will
amplify that signal, and create another amplified tone at a
frequency of $f_{i}=2f_p-f_s$, usually referred to as the idler or
intermodulation tone. As we increase the pump in order to obtain
higher gains, the optimal pump frequency will shift down as a
consequence of the Kerr-nonlinearity of the cavity.



\section{Results and Analysis}

\begin{figure}
\includegraphics{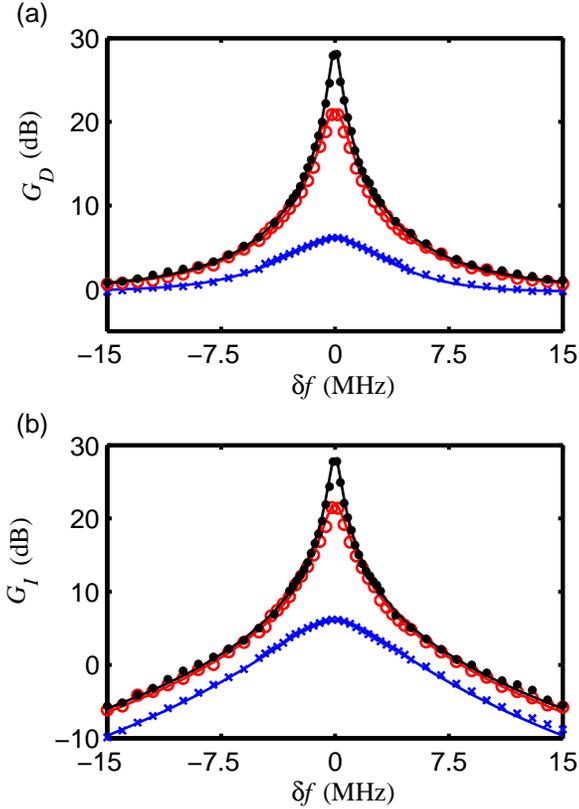}
\caption{\label{fig:Fig3} Gain measurements. (a) and (b) Direct and
and intermodulation gain as functions of signal-pump detuning
(points) and predictions of
 \cite{yurke2006} (lines) for three different pump powers and frequencies: $P=0.95\ P_c$ and $f_p=6.176$~GHz (dots), $P=0.9\ P_c$ and
$f_p=6.1768$~GHz (circles), and for $P=0.6\ P_c$ and
$f_p=6.1806$~GHz (crosses). For each pump power, we find an optimum
pump frequency. The optimum frequency decreases with increasing pump
power due to the Duffing-like behavior of the resonator
\cite{yurke2006}. Then we sweep the signal frequency over a range
around the optimum pump frequency. The signal power used to measure
the gains was -160 dBm.}
\end{figure}

We study the amplifier's properties using the apparatus shown in
Fig.~\ref{fig:Fig1}a. As described in \cite{Castellanos}, we
determine the critical pump power at which the JPA gain diverges. We
then characterize the intermodulation gain $G_I$ and direct gain
$G_D$ for pump powers less than this critical power. We define $G_D$
as the ratio between the incident and reflected signal power; $G_I$
is the ratio between the intermodulation tone and the incident
signal power. To characterize the bandwidth of the parametric
amplifier, we measure the frequency dependance of both gains. For
sufficiently large gains, these are Lorentzian functions of the
detuning $\delta f$ between the signal and pump frequencies. For
each pump power we determine the 3 dB bandwidth $\Delta f_{1/2}$ and
gain at $\delta f=0$. We observe that with increasing gain, $\Delta
f_{1/2}$  is reduced. We find that the expression $\sqrt{G_D}\Delta
f_{1/2}\approx \gamma$ is correct, within a factor of two
\cite{yurke2006}, where $\gamma$ is the half width at half-maximum
of the resonator, and is given by $\gamma=
\gamma_{c1}+\gamma_{c2}+\gamma_i$ . This expression is also correct
for $G_I$ if $G_I>4$. These results are shown in figs.
\ref{fig:Fig3}(a) and \ref{fig:Fig3}(b), where we plot both $G_D$
and $G_I$ as a function $\delta$$f$ for three different pump powers
for $f_{res}=6.184$~GHz ($\Phi=0.37\Phi_0$). Qualitatively, this
behavior is expected for parametric amplifiers, and we find
quantitative agreement with the theory of \cite{yurke2006}. From
this plot, we estimate the $3$~dB bandwidth to be about $1.38$~MHz
when $G_D$ and $G_I$ are $22$~dB. Although we can observe some
internal loss, this loss won't significantly affect the gain or the
bandwidth of the amplifier as long as $\gamma_i < \gamma_{c1}
+\gamma_{c2}$. However, any loss will lead to some additional noise
added by the amplifier, a quantity which we recently measured
\cite{Castellanos2}.

\begin{figure}
\includegraphics{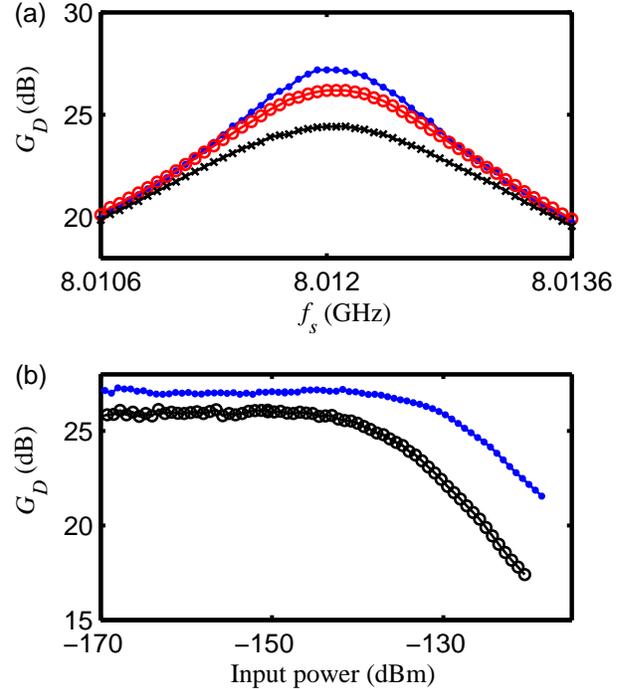}
\caption{\label{fig:Fig4} (a) Direct gain as a function of signal
frequency for $\Phi=0.15\Phi_0$ for three different signal powers,
and $f_p=8.0121$ GHz, showing the expected saturation behavior at
large enough signal powers.  Signal powers shown are -145 dBm
(dots), -130 dBm (circles) and -125 dBm (crosses). (b) Direct gain
as a function of input signal powers for $f_{p}=6.1766$ GHz
(circles) and $f_{p}=8.0121$ GHz (dots). The signals were applied
60~kHz above the pump frequency.}
\end{figure}

For low enough powers in the signal, the gain of the amplifier will
be linear, i.e., the output power of the amplifier will depend
linearly with the input power. However, as we increase the signal
power we will start saturating the JPA, as shown in Fig.
\ref{fig:Fig4}a. We have estimated saturation power and the dynamic
range of this amplifier at 2 different resonant frequencies. We
determine the saturation power as the input power for which the gain
is reduced by 1 dB (Fig. \ref{fig:Fig4}b). In Fig. \ref{fig:Fig4}b
we plot $G_D$ as function of input signal power for the two cavity
resonance frequencies. At 8.012~GHz with a pump power of $P=-76$ dBm
and $G_D=27$~dB, the 1~dB compression point is $-130$~dBm. From
noise measurements described in \cite{Castellanos2}, the minimum
detectable signal for this amplifier in 1 Hz band is $0.77 \hbar
\omega$. From these two measurements, we estimate the dynamic range
to be 74 dB/Hz. At 6.177 GHz with a pump power of P$=-84$~dBm and
$G_D=26$~dB, the 1~dB compression point is $-137$~dBm and the
dynamic range is $67$~dB/Hz. The reduction in dynamic range is
expected: when we tune down the resonance frequency by applying a
magnetic field, we are reducing the critical currents of the SQUIDs
and the critical power of the resonator.

As a rule of thumb, we have observed that the saturation powers for
large gains can be found as $P_{sat} \approx
(\gamma_{c1}/\gamma_{c2})P_c/(20G_D)$, where $P_c$ is the critical
pump power. The factor $\gamma_{c1}/\gamma_{c2}$ arises because we
applied the pump to the weakly coupled port, but applied the signal
to the strongly coupled port. This expression is reasonable as we
expect to observe saturation behavior when the amplified signal
power becomes a significant fraction of the pump power inside the
cavity.

\section{Conclusion}

Applications for an amplifier of this type would include amplifying
microwave signals that encode the motion of a nanomechanical
oscillator or a superconducting qubit. It is clear that for largest
bandwidth one would like to operate with the minimum gain necessary.
We can make an estimate of the minimum gain by asking what JPA gain
would make the noise added by the following amplifier equal to the
vacuum noise. If the following amplifier is a state-of-the-art HEMT
amplifier operating at 5 GHz and assuming a noise temperature of 5
K, we would require 16 dB of gain. For this realization of a JPA,
the bandwidth would then be approximately 2 MHz. This bandwidth is
well suited for detecting nanomechanical oscillators of the type in
\cite{Regal2008}, but marginal for measuring the state of a
superconducting qubit \cite{Houck}. However, it is plausible to
increase the gain-bandwidth product of the JPA by implementing
larger coupling capacitors, hence increasing the cavity linewidth.

\section*{Acknowledgment}

The authors acknowledge support from the National Institute of
Standards and Technology (NIST), from the National Science
Foundation, and from a NIST-University of Colorado seed grant. We
thank S.~M. Girvin and J. D. Teufel for valuable conversations and
technical assistance. K.~W. Lehnert is a member of NIST's Quantum
Physics Division.





\end{document}